\documentclass{article}

\PassOptionsToPackage{numbers, compress}{natbib}
\usepackage[preprint]{neurips_2023}

\usepackage[utf8]{inputenc} 
\usepackage[T1]{fontenc}    
\usepackage{hyperref}       
\usepackage{url}            
\usepackage{booktabs}       
\usepackage{amsfonts}       
\usepackage{nicefrac}       
\usepackage{microtype}      
\usepackage{xcolor}         
\usepackage{amsmath}
\usepackage{amssymb}
\usepackage{mathtools}
\usepackage{amsthm}
\usepackage{multirow}
\usepackage{wrapfig}
\usepackage[capitalize,noabbrev]{cleveref}
\usepackage{graphicx}
\usepackage{subfigure} 
\usepackage{float}
\usepackage{algorithm} 
\usepackage{algorithmic}
\usepackage{textcomp} 
\usepackage{listings} 
\usepackage{enumitem}
\usepackage{utfsym}
\usepackage{times}
\usepackage{latexsym}
\usepackage{fancyvrb}
\usepackage{CJKutf8}

\newcommand{\commentout}[1]{}
\newcommand{\chn}[1]{\begin{CJK*}{UTF8}{gbsn}{#1}\end{CJK*}}

\title{FireRedTTS: A Foundation Text-To-Speech Framework for Industry-Level Generative Speech Applications}

\author{
FireRed Team\thanks{Authors (alphabetical order): Hao-Han Guo,  Yao Hu, Kun Liu, Fei-Yu Shen, Xu Tang, Yi-Chen Wu, Feng-Long Xie, Kun Xie, Kai-Tuo Xu. Corresponding author: Feng-Long Xie (xiefenglong7825@gmail.com)} \\
Xiaohongshu
}

\begin{document}
\maketitle
\begin{abstract}
This work proposes FireRedTTS, a foundation text-to-speech framework, to meet the growing demands for personalized and diverse generative speech applications. The framework comprises three parts: data processing, foundation system, and downstream applications. First, we comprehensively present our data processing pipeline, which transforms massive raw audio into a large-scale high-quality TTS dataset with rich annotations and a wide coverage of content, speaking style, and timbre. Then, we propose a language-model-based foundation TTS system. The speech signal is compressed into discrete semantic tokens via a semantic-aware speech tokenizer, and can be generated by a language model from the prompt text and audio. Then, a two-stage waveform generator is proposed to decode them to the high-fidelity waveform. We present two applications of this system: voice cloning for dubbing and human-like speech generation for chatbots. The experimental results demonstrate the solid in-context learning capability of FireRedTTS, which can stably synthesize high-quality speech consistent with the prompt text and audio. For dubbing, FireRedTTS can clone target voices in a zero-shot way for the UGC scenario and adapt to studio-level expressive voice characters in the PUGC scenario via few-shot fine-tuning with 1-hour recording. Moreover, FireRedTTS achieves controllable human-like speech generation in a casual style with paralinguistic behaviors and emotions via instruction tuning, to better serve spoken chatbots. Our demos are available at \url{https://fireredteam.github.io/demos/firered_tts}.
\end{abstract}


\section{Introduction}
\label{sec:intro}


Text-to-speech synthesis (TTS) has been playing a critical role in intelligent interaction \cite{guo2021conversational, budzianowski2024pheme}, e.g., virtual assistants, chatbots, and AI content creation, e.g. video dubbing \cite{hu2021neural, liu2024m3tts}. With the development and popularization of these AI products, TTS is facing a new challenge: providing personalized and diverse speech generation, to better satisfy users' variant requirements on AI products. 

This requirement calls for a powerful model to understand speech and generate arbitrary speech. Recently, the success of large language models (LLMs) \cite{touvron2023llama2,gpt4,gozalo2023chatgpt} demonstrates their great capability in sequential modeling, implying their potential in speech applications. The large language models trained with massive speech data have shown impressive performance in speech generation, such as VALL-E \cite{VALLEX}, TorToiSe \cite{tortoise}, BASE-TTS \cite{lajszczak2024base}, Seed-TTS \cite{anastassiou2024seed}, CosyVoice \cite{du2024cosyvoice}, etc. They can produce expressive and diverse speech via zero-shot in-context learning \cite{dong2022survey}. Inspired by them, we propose a language-model-based foundation TTS framework, FireRedTTS, to better support industry-level generative speech applications. 

The proposed framework comprises three parts: data processing, foundation system, and downstream applications. First, we present a complete and effective data processing pipeline to create a high-quality large-scale TTS dataset from massive raw audio data in five steps: speech enhancement, speech segmentation, speaker clustering, transcribing, and data filtering. Then, we employ this dataset to train the proposed language-model-based foundation TTS system. In this system, we propose a semantic-aware speech tokenizer to compress the long speech sequence into discrete speech tokens with sufficient semantic information. These tokens can be generated from a text-to-speech language model, and reconstructed to the audio by a two-stage token-to-waveform generator, i.e. first converting semantic tokens into the Mel spectrogram via a Mel decoder, and then generating the audio with a high sampling rate of 48 kHz via a super-resolution neural vocoder.

We present two downstream applications of the proposed system: voice cloning for dubbing and human-like speech generation for chatbots. Compared to the conventional TTS approach\cite{tacotron2,fastspeech2}, we achieve studio-level speech synthesis with much less training data, e.g. zero-shot voice cloning for non-studio user-generated content (UGC) scenario and few-shot speaker adaptation with 1 hour for professional user-generated content (PUGC) scenario, making video dubbing customizable. Moreover, we achieve human-like speech generation via instruction tuning, enabling the chatbot to speak in a casual style with multiple emotions and rich paralinguistic behaviors.

The contributions of our work are summarized as follows:
\begin{itemize}
    \item We propose a complete and detailed data processing pipeline to forge the raw, noisy audio dataset into a clean TTS dataset with rich annotations and high coverage of content, speaking style, and timbre for foundation TTS training.
    \item We propose a foundation TTS system based on the language model. Specifically, we propose a semantic-aware tokenizer to effectively convert speech signals into discrete semantic tokens, which can be generated from the prompt text and audio via a text-to-speech language model. Moreover, the two-stage token-to-waveform generation is proposed to achieve high-fidelity audio synthesis.    
    \item We present two applications of FireRedTTS, voice cloning for dubbing and human-like speech generation for chatbots, demonstrating the potential of FireRedTTS in downstream speech applications.
\end{itemize}

\section{Data Processing Pipeline}
\label{sec:data_processing_pipeline}







\begin{figure*}[!ht]
\centering
\includegraphics[width=\linewidth]{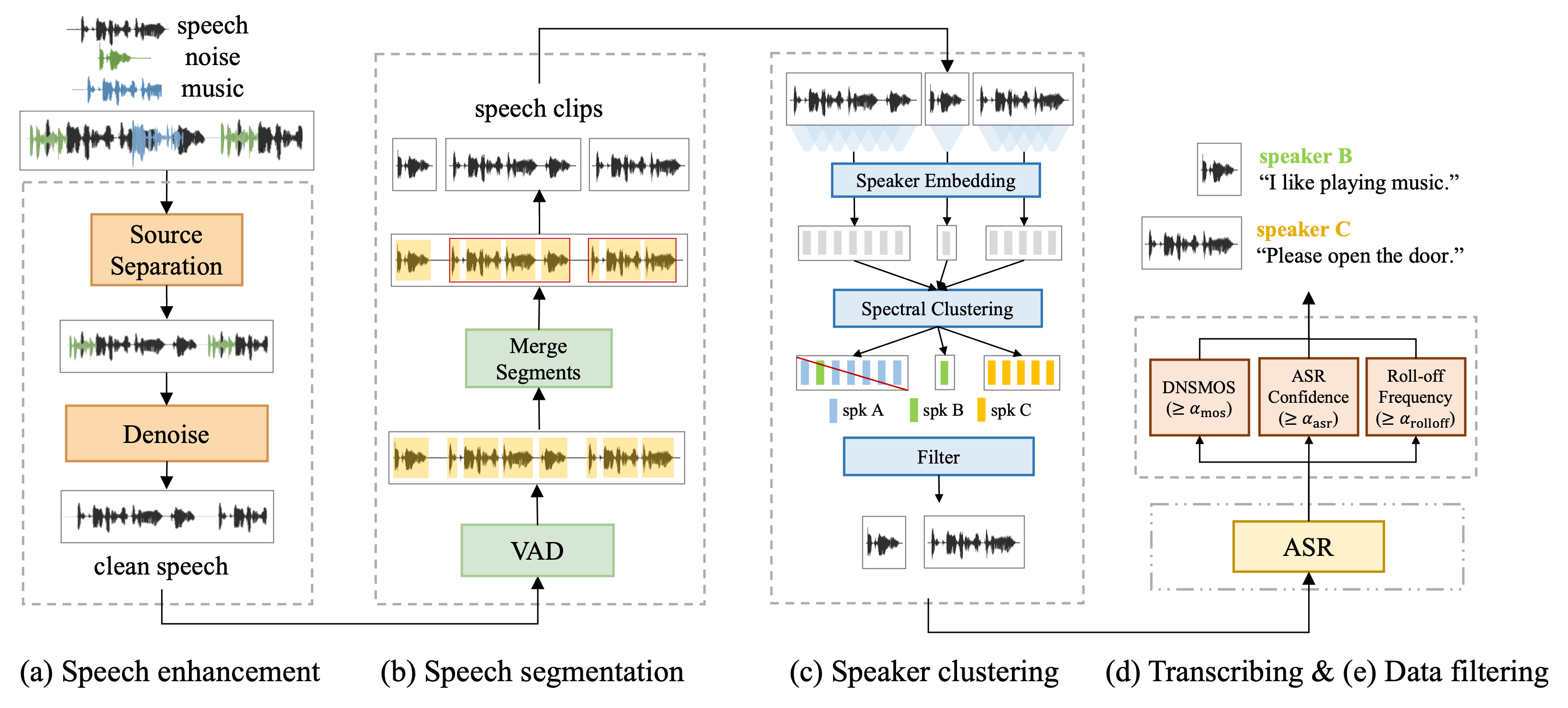}
\caption{Overview of the data processing pipeline. The roll-off frequency represents the frequency below which 99.5\% of the total spectral energy is contained. The thresholds for $\alpha_{\text{mos}}$, $\alpha_{\text{rolloff}}$, and $\alpha_{\text{asr}}$ are set to 3.3, 7kHz, and 0.8, respectively.}
\label{img:pipeline_overview}
\end{figure*}

As shown in Figure \ref{img:pipeline_overview}, we propose a data processing pipeline to create a large-scale TTS dataset from massive in-the-wild speech data. This pipeline comprises five steps: speech enhancement, speech segmentation, speaker clustering, transcribing, and data filtering, which will be elaborated in the section.

\noindent\textbf{Speech Enhancement}: First, speech enhancement removes non-speech components in the audio, e.g., background music and noise, to provide clean speech audio. This module is composed of two models: a music source separation model to remove the background music \cite{rouard2023hybrid} and a speech enhancement model \cite{schroter2022deepfilternet} for speech denoising. 

\noindent\textbf{Speech Segmentation}: We then perform speech segmentation to cut the long speech audio into short segments ranging from 2s to 20s to adapt TTS training. We train a frame-wise TDNN-based voice activity detection (VAD) model \cite{bai2019voice} to accurately detect speech segments from the Mel spectrogram with a frameshift of 25ms. We combine adjacent segments with interval silence shorter than 1 second and extend the segment boundary with 0.3 seconds to avoid cutoffs in aspiration and trailing.

\noindent\textbf{Speaker Clustering}: Due to the lack of speaker information, we perform spectral clustering on speaker embeddings \cite{wang2023wespeaker} of all speech segments. Specifically, we conduct K-Means clustering and iteratively merge clusters with cosine similarity higher than 0.8. Finally, each chunk of the segment is attributed with a speaker centroid and the corresponding speaker ID. A segment with multiple IDs will be denoted as a multi-speaker segment and removed from the dataset. Besides, we notice that a segment too far away from its corresponding speaker centroid usually implies lower speech quality or a mixture of speakers, which is also filtered out. 

\begin{figure}[htp]
\centering
\includegraphics[width=0.5\linewidth]{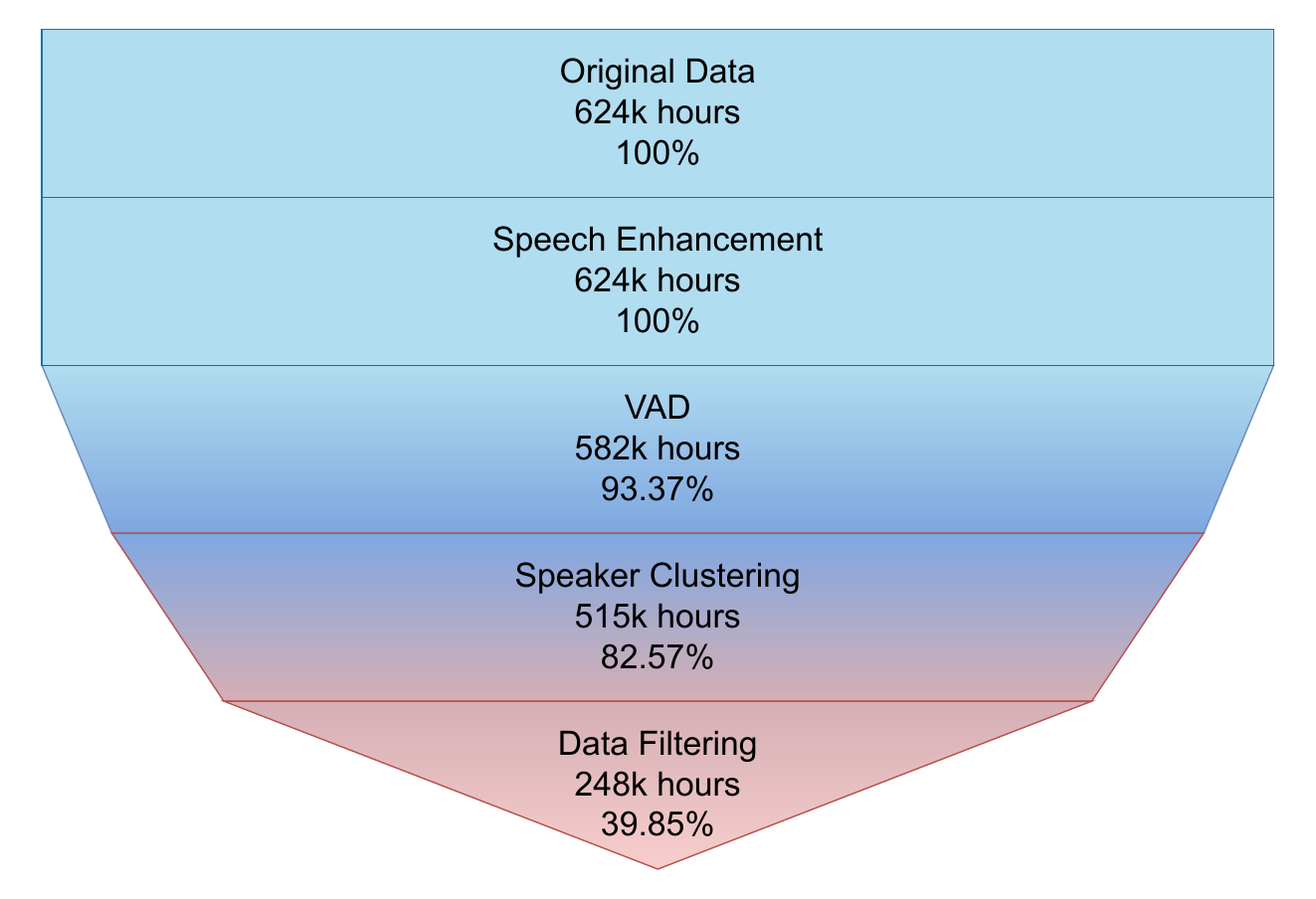}
\caption{The amount of speech data remained after each processing step.}
\label{img:data_funnel}
\end{figure}

\noindent\textbf{Transcribing}: We employ a two-pass Transducer-based end-to-end ASR model \cite{sainath2019two} to transcribe speech segments. The batched Transducer beam search is applied for inference to transcribe massive data efficiently.

\noindent\textbf{Data Filtering}: Finally, we perform data filtering to remove all unqualified speech segments after the long chain of speech pre-processing. The filtering is conducted from three aspects:
\begin{itemize}
    \item Speech quality: we estimate the overall speech quality using DNSMOS \cite{reddy2022dnsmos}, and keep only speech segments with the DNSMOS P.835 OVRL score \cite{yu2024autoprep} higher than 3.3. 
    \item Sampling rate: we aim to create a high-sampling-rate speech dataset. Hence, we filter low-sampling-rate (e.g., 8 kHz) audio saved in the high-sampling-rate format. We calculate the roll-off frequency whose energy under it holds 99.5\% of the total energy. Then, only speech segments with roll-off frequency higher than 7 kHz are kept in the dataset.
    \item Transcription confidence: We discard speech segments with ASR confidence scores below 0.8 to mitigate the impact of transcription errors on TTS training.
\end{itemize}

Finally, we collected around 248k hours of labeled speech data from the 624k hours of unlabeled audio via the proposed pipeline. The funnel chart shown in Figure \ref{img:data_funnel} presents a detailed analysis of how much data is filtered in each step. In this work, we use a subset with 150k hours (110k hours for Chinese and 40k hours for English) for model training.

\section{Foundation TTS System}

\begin{figure}[htp]
\centering
\includegraphics[width=\linewidth]{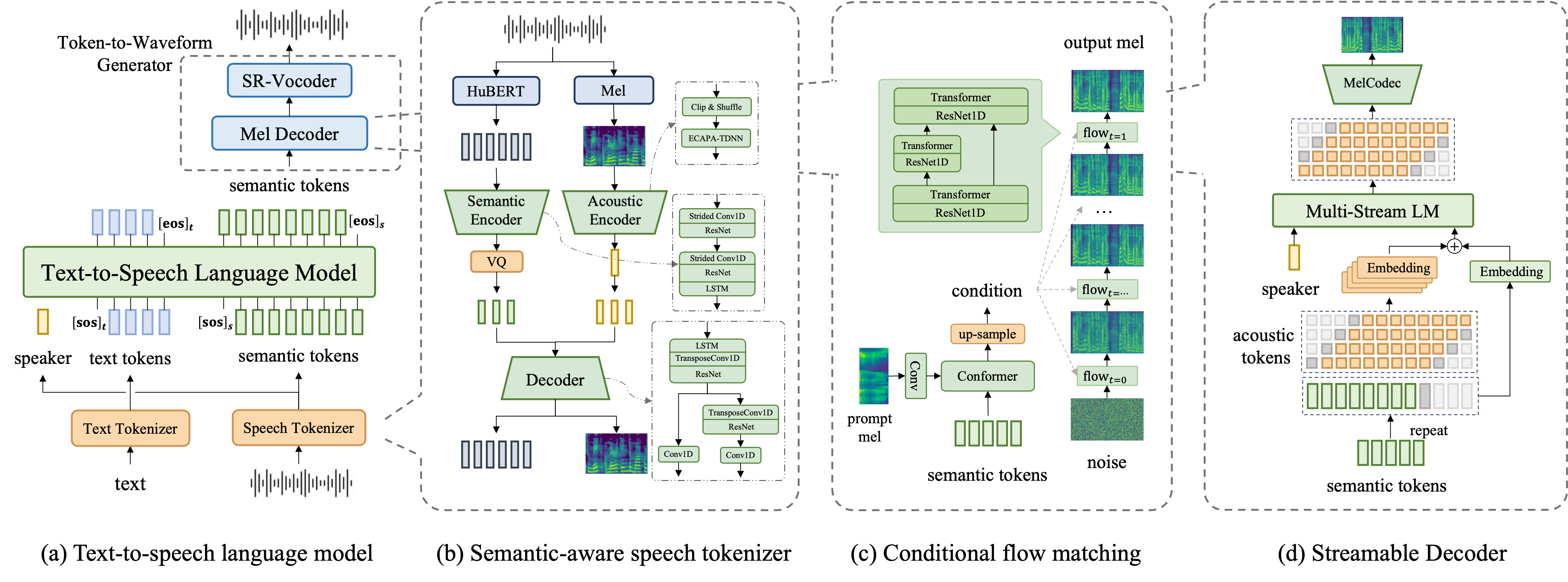}
\caption{An overview of the FireRedTTS foundation system with a text-to-speech language model that maps text tokens to semantic tokens, and a two-stage token-to-waveform generator. The semantic tokens are extracted using a semantic-aware speech tokenizer. In Mel decoder, (c) illustrates a flow-matching-based approach that transforms the noise into a spectrogram using semantic tokens as conditions, and (d) depicts a streamable decoder based on a multi-stream language model and a Mel Codec.}
\label{img:tts_model_framework}
\end{figure}

As shown in Figure \ref{img:tts_model_framework}, we propose a language-model-based foundation TTS system. This system is composed of three modules: speech tokenizer, text-to-speech language model, and token-to-waveform generator.

\subsection{Semantic-Aware Speech Tokenizer}

As shown in Figure \ref{img:tts_model_framework}(b), the semantic-aware speech tokenizer (SAST) aims to compress speech signals into discrete tokens for TTS modeling. It is composed of the semantic encoder, the acoustic encoder, and the decoder.

\noindent\textbf{Semantic Encoder}: This module aims to transform the speech signal into a sequence of semantic tokens. It first utilizes a pre-trained self-supervised learning (SSL) model, HuBERT \cite{hsu2021hubert}, to translate speech signals into a semantic embedding sequence. Then, the semantic encoder, constructed with ResNet architecture, is employed to process and down-sample the sequence further into an encoding sequence. Finally, we follow \cite{guo2024addressing} to quantize this sequence into the discrete sequence with a frameshift of 40ms and a codebook of 16,384 codewords.

\noindent\textbf{Acoustic Encoder}: Meanwhile, we employ an ECAPA-TDNN-based acoustic encoder, as delineated in \cite{dawalatabad2021ecapa}, to derive a global embedding at the utterance level from the speech signals. This captures critical time-invariant characteristics, including speaker identity, speaking style, and the acoustic environment. This embedding can be utilized for the emulation of a target voice in zero-shot TTS scenarios. To ensure that content information is not leaked to this embedding, we apply a straightforward pre-processing strategy, "Clip\&Shuffle", to the Mel spectrogram to remove short-time variant information. The process first selects a segment constituting 25\% to 75\% of the total utterance duration, which is then partitioned into 1-second slices. These slices are subsequently rearranged in a random sequence to obtain the global embedding via the acoustic encoder.

\noindent\textbf{Decoder}: In the decoder, the global embedding is duplicated and added with the quantized sequence to form the decoder input. It is then processed by ResNet blocks with the transposed convolutional layers for up-sampling to reconstruct both SSL features and acoustic features. 

SAST is trained with the following loss function:
\begin{align}
    \mathcal{L}_c = \lambda_{vq} * \mathcal{L}_{vq} + \lambda_{s} * \mathcal{L}_{s} + \lambda_a * \mathcal{L}_{a}
\end{align}
where $\lambda_{vq}, \lambda_s, \lambda_a$ are weight coefficients. $\mathcal{L}_{vq}$ is the VQ loss, i.e. L2 loss between embeddings before and after the quantization \cite{vqvae}. 
$\mathcal{L}_{s}$ is the L2 loss between the SSL features and the reconstructed ones, and $\mathcal{L}_{a}$ is the L2 loss between the ground-truth Mel spectrogram and the reconstructed one. In this work, we adopt $\lambda_{vq} = 1, \lambda_{s} = 1000, \lambda_a = 1$, and train the model with a large batch size of 6400 seconds for 300k iterations.


\subsection{Text-To-Speech Language Model}

Inspired by LLM successes, we formulate TTS as a next-token prediction task using a decoder-only autoregressive Transformer. As shown in Figure \ref{img:tts_model_framework} (a), we aim to predict semantic tokens from the prompt text and audio. The text is encoded into the token sequence by a BPE-based text tokenizer \cite{whisper}. The prompt audio is embedded into an utterance-level embedding via the acoustic encoder of SAST. In training, we extract both speech tokens and speaker embedding from the target audio and embed them and the text, respectively, which are then concatenated together as the input of the GPT-like decoder-only Transformer. \cite{brown2020language}. During inference, we can autoregressively generate semantic tokens via in-context learning, i.e. giving the prompt text and the speaker embedding from the prompt audio. In this work, we train a 30-layer decoder-only Transformer (400M) with a feature dimension of 1024.

\subsection{Token-to-Waveform Generation}

This work aims to synthesize diverse speech audio, including high-fidelity samples with a sampling rate of 48 kHz. However, our training dataset predominantly comprises low-sampling-rate ($\leq$ 24 kHz) audio, which cannot be directly utilized for training a high-sampling-rate waveform generator. To address this limitation, we propose a novel two-stage token-to-waveform generator. The first stage involves training a Mel decoder to convert semantic tokens into Mel spectrograms extracted from low-sampling-rate (16 kHz) audio. Subsequently, we employ a super-resolution neural vocoder \cite{liu2024audiosr} to generate high-sampling-rate (48 kHz) audio from the Mel spectrogram. This approach enables the system to effectively leverage the entire training dataset. For the token-to-Mel generation process, we introduce two decoders: a flow-matching-based decoder and a streamable decoder.

\subsubsection{Flow-Matching}

To enhance the quality of generated Mel spectrograms, we employ the flow-matching structure \cite{mehta2024matcha, lipman2022flow, du2024cosyvoice} that learns the denoising process from noise into the spectrogram distribution. At inference, it can iteratively convert the Gaussian noise into a high-quality Mel spectrogram with semantic tokens as input conditions.

Specifically, flow matching learns the time-conditioned transformation $\phi_t$ into the spectrogram sample $x_1$, starting from a noisy sample $x_0$ drawn from standard Gaussian distribution. This transformation is governed by an ordinary differential equation (ODE):
\begin{align}
\frac{d}{dt}\phi_t(x) = v_t(\phi_t(x))
\end{align}
The time-conditioned vector field $v_t$ can be chosen as the optimal transport path:
\begin{align}
\phi_t^{\text{OT}}(x_0, x_1) = (1-(1-\sigma)t)x_0 + tx_1 
\end{align}
\begin{align}
v_t^{\text{OT}} = \frac{d}{dt}\phi_t^{\text{OT}}(x_0, x_1) = x_1 - (1-\sigma)x_0
\end{align}
where $\sigma$ is a small constant. A neural network, parameterized by $\theta$, estimates the vector field. As illustrated in Figure \ref{img:tts_model_framework}(c), semantic tokens are first encoded by a Conformer and then upsampled to match the length of the Mel spectrogram. The encoded feature (denoted as condition $\Psi$), together with the sample $x_t$ and timestep $t$, are fed into a U-Net-based estimator for vector field prediction.
\begin{align}
v_t^{\text{pred}} = \text{NN}_{\theta}(x_t, t, \Psi)
\end{align}
Since semantic tokens contain limited timbre information, we propose introducing a cross-attention layer after each self-attention layer in Conformer to extract timbre from the Mel spectrogram of the reference audio, i.e. another audio of the same speaker as the target audio. 

The training objective is defined as:
\begin{align}
\mathcal{L}_{fm} = \left \| v_t^{\text{pred}} - v_t^{\text{OT}} \right \|^2
\end{align}
Moreover, to better generate the Mel spectrogram, we apply the Classifier-Free Guidance (CFG \cite{ho2022classifier}) technique. In training, we randomly drop conditions in one batch with a probability of 20\%. At inference, the vector field is estimated with both conditioned and unconditioned predictions:
\begin{align}
v_t^{\text{cfg}} = (1+\alpha)\text{NN}_{\theta}(x_t, t, \Psi) - \alpha\text{NN}_{\theta}(x_t, t)
\end{align}
where $\alpha$ is 0.7.

\subsubsection{Streamable Decoder}

Flow matching presents impressive generative quality in Mel spectrogram generation, but has been troubled by the slow iterative inference, making it challenging to implement streaming generation. To avoid this issue, we propose a streamable decoder incorporating a Mel Codec and a multi-stream language model.

As shown in Figure \ref{img:tts_model_framework}(d), we first follow \cite{socodec} to train a streamable CNN-based GAN-based Mel codec to learn a multi-stream discrete representation, i.e. each frame consisting of multiple tokens to preserve sufficient acoustic details. Specifically, the Mel spectrogram with a frameshift of 10ms is compressed into a four-stream discrete sequence with a frameshift of 20ms by four codebooks with 16,384 codewords respectively. This sequence can be reconstructed back via the Mel codec decoder, and predicted from semantic tokens via a multi-stream LM \cite{copet2024simple} with a ``delay pattern''.

The acoustic LM combines the semantic sequence and the acoustic sequence with a ``lookahead pattern''. Specifically, we will first embed and upsample the semantic sequence by repetition to align to the acoustic sequence, and then add them with a delay, i.e. $s_{i-d} + a_{i}$, where $a_{i}$ is the $i$-th acoustic embedding, $s_{i-d}$ is the $(i-d)$-th semantic embedding, and $d$ is the number of lookahead frames. This approach enables a synchronous streaming process: once we receive semantic tokens from the TTS language model, we start to generate acoustic tokens and the corresponding waveform in streaming synchronously.

in streaming.

\subsubsection{Super-Resolution Vocoder}

To train the super-resolution vocoder, we employ a carefully curated subset of the original training corpus, comprising 294 hours of high-sampling-rate audio. From these files, we extract Mel spectrograms with a frameshift of 10 ms by downsampling the audio to 16 kHz. This refined dataset serves for training a BigVGAN-V2-based neural vocoder \cite{bigvgan}. The vocoder upsamples the Mel spectrogram by a factor of 480, effectively transforming it into a high-fidelity waveform with a sampling rate of 48 kHz. This two-stage approach enables the generation of high-quality, high-sampling-rate audio while efficiently utilizing the predominantly low-sampling-rate training data.

\section{Downstream TTS Application}

The foundation system is trained on a large-scale speech corpus to capture diverse speaker identities and speaking styles. This comprehensive training enables the system to adapt to a wide range of scenarios. Consequently, in practical applications, we can leverage this foundation system for downstream tasks with minimal data requirements. In this work, we highlight two pivotal applications that demonstrate the system's versatility and efficiency: voice cloning for dubbing and human-like speech generation for chatbots.

\subsection{Voice Cloning for Dubbing}
\label{ss:clone}

Voice cloning technology has found widespread application in dubbing for video editing and content creation, enabling the synthesis of speech audio with custom text in a target voice. In real-world applications, voice-cloning-based dubbing primarily serves two distinct scenarios: user-generated content (UGC) scenarios and professional user-generated content (PUGC) scenarios. 

UGC scenarios typically present a data-limited scenario, necessitating the synthesis of target voices using only a few seconds or minutes of low-quality (non-studio) recordings as references. In these cases, we employ in-context learning to achieve text-to-speech (TTS) in a zero-shot (or one-shot) manner by utilizing speaker embeddings extracted from prompt audio. The efficacy of this approach hinges on obtaining a global embedding that accurately represents the target voice. However, in practice, the low-quality, noisy recordings commonly uploaded by users frequently interfere with the extraction of precise speaker embeddings, thereby degrading voice cloning performance. To mitigate this issue, we employ ``prompt processing", i.e. speech enhancement on the prompt audio, to derive clean reference audio. This crucial step enhances the quality of speaker embeddings and, consequently, improves the overall performance of voice cloning in UGC scenarios.

PUGC scenarios typically provide studio-quality audio data featuring distinctive voices performed by professional voice actors. The high expressiveness inherent in such data presents a significant modeling challenge. To address this, we propose adapting our foundation model to the target voice through supervised fine-tuning. This approach enables us to quickly capture the rich, nuanced characteristics of professional voice recordings by leveraging the generalization capability of our pre-trained foundation model, achieving high-quality synthesis with less target data.

\begin{figure}[htp]
\centering
\includegraphics[width=0.6\linewidth]{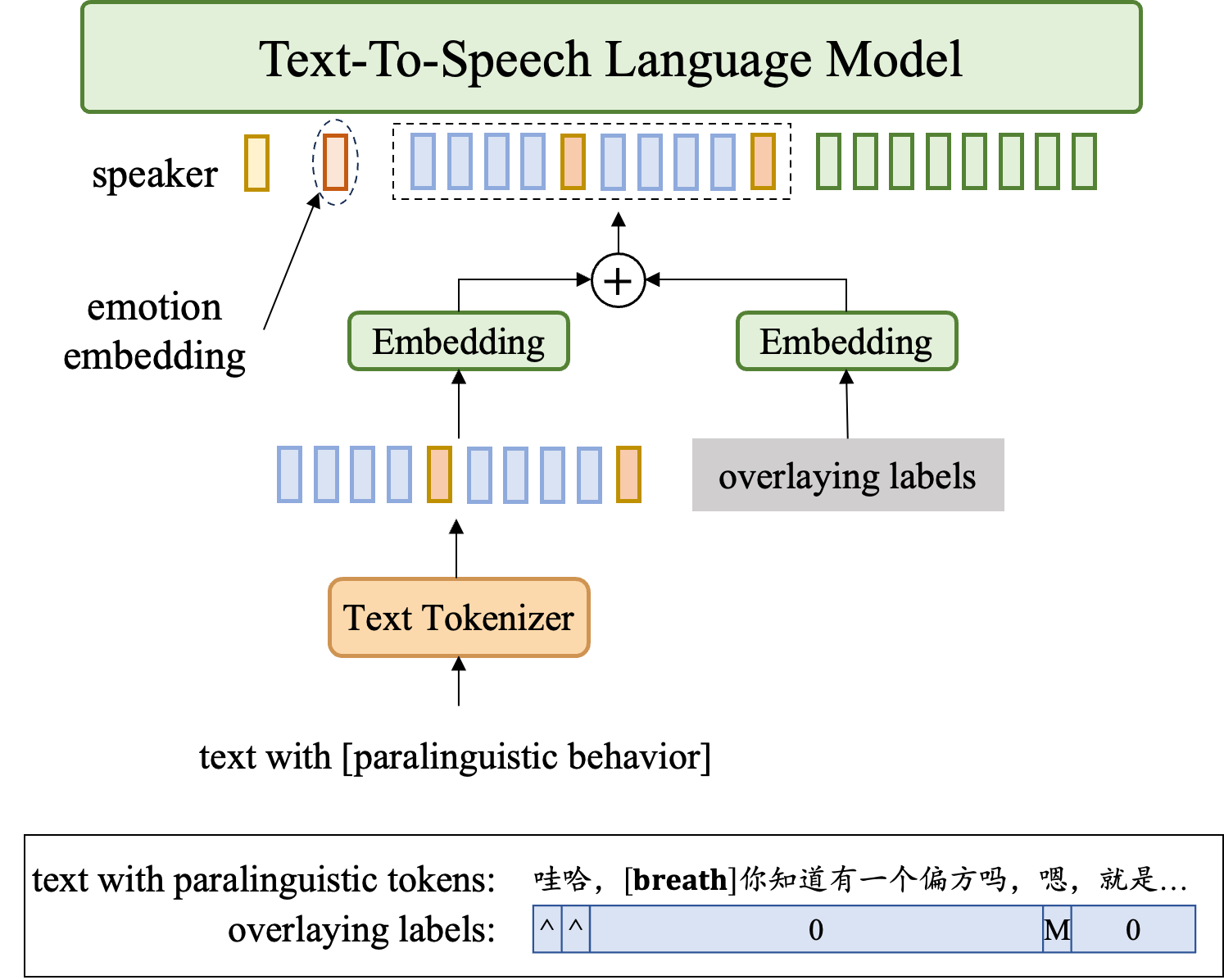}
\caption{Instruction tuning for human-like speech generation.}
\label{img:spon_adapter}
\end{figure}

\subsection{Human-like Speech Generation for Chatbots}

The proliferation of chatbots has created an urgent demand for natural and human-like speech generation to achieve immersive human-machine interaction. While our foundation system can already synthesize natural speech with arbitrary speaking styles, it still lacks fine-grained controllability in human-like speech generation, particularly in conveying emotions and paralinguistic behaviors. To address this limitation, we present an instruction-tuning-based approach utilizing a specialized 50-hour dataset rich in emotional and paralinguistic content.

As illustrated in Fig. \ref{img:spon_adapter}, we introduce emotions and paralinguistic behaviors to the LM by modifying the prompting sequence. Initially, we derive an emotional embedding from a dedicated embedding layer encompassing four distinct categories: neutral, happy, sad, and angry. This emotional embedding is then inserted into the prompting sequence. Notably, during the training phase, we extract the speaker embedding from a separate audio sample of the same speaker. This deliberate approach serves to disentangle speaker and emotion representations.

Paralinguistic behaviors are incorporated into the text sequence through two distinct modes: token insertion and embedding injection. For acoustic-event-based behaviors, such as pauses, elongations, repetitions, laughter, and breathing sounds, we insert pre-designed tokens directly into the raw text sequence. Conversely, for overlapped paralinguistic labels, e.g. ``speaking while laughing" or ``emphasis", we introduce a dedicated paralinguistic behavior embedding layer, allowing these labels to be associated with specific words. This approach enables us to achieve controllable paralinguistic behavior synthesis. In the current implementation, we have successfully incorporated 13 paralinguistic behaviors, as detailed in Table \ref{tab:data_finetune_spon}. This comprehensive set of behaviors allows for a human-like speech synthesis system, capable of mimicking the complex vocal patterns observed in natural human conversations.

\begin{table}[!ht]
\centering
\renewcommand{\arraystretch}{1.1}
\begin{tabular}{ccccc}
\toprule
\textbf{\begin{tabular}[c]{@{}c@{}}Para-Linguistic Behaviors\end{tabular}} & \textbf{\begin{tabular}[c]{@{}c@{}}Text Input Format\end{tabular}} & \textbf{\begin{tabular}[c]{@{}c@{}}Label Type \end{tabular}} \\ \midrule
Char repetition & \chn{我} $[\text{hic}]$ \chn{我} & token insertion \\
Word repetition & \chn{就是} $[\text{rep}]$ \chn{就是} & token insertion \\
Elongation & \chn{就是} $[\text{elong}]$ & token insertion \\
Hissing & $[\text{sss}]$ & token insertion \\
Dental click & $[\text{tsk}]$ & token insertion \\
Breath & $[\text{breath}]$ & token insertion \\
Laugh & $[\text{laugh}]$ & token insertion \\ \midrule
Speak with a laugh & \chn{真是笑}\^{}\chn{死}\^{}\chn{我}\^{}\chn{了}\^{} & embedding injection \\
Emphasis & \chn{你真}${}^{@}$\chn{棒}${}^{@}$ & embedding injection \\ \midrule
Filled pause & \chn{嗯}${}^{P}$ & embedding injection \\
Confirmation & \chn{啊}${}^{C}$ & embedding injection \\
Realization & \chn{哦}${}^{R}$ & embedding injection \\
Surprise & \chn{哦}${}^{S}$ & embedding injection \\

\bottomrule
\end{tabular}
\caption{The list of para-linguistic behaviors implemented in our work.}
\label{tab:data_finetune_spon}
\end{table}





\section{Results}

\subsection{Foundation Model Evaluation}


\subsubsection{Consistency Evaluation}
\label{ssec:comos}

The proposed foundation system is expected to synthesize speech via in-context learning, presenting the content of the prompt text with the timbre and speaking style of the prompt audio. Hence, to measure the consistency between the synthesized speech and prompts, we conduct a consistency MOS (CoMOS) test. We create a test set with 94 Chinese <text, audio> pairs, covering various emotions and speaking styles. Each sample consists of two consecutive segments, with the first segment serving as the prompt audio and the second as the synthesis target. During the evaluation, each listener is asked to rate the synthesized audio on a scale of 1 to 5, based on the consistency between synthesized audio and prompts in content, speaking style, and timbre.

As illustrated in Table \ref{tab:exp_content_consistency}, we compare three groups of audio: the ground-truth audio, the synthesized audio from CosyVoice \cite{du2024cosyvoice}, and FireRedTTS. First, the ground-truth audio achieves the highest CoMOS score of 4.53. CosyVoice, which is a high-quality LM-based large-scale TTS system trained on over 100,000 hours of data, exhibits a high synthesis quality with a CoMOS score of 4.15. However, there is still a significant gap compared to the ground-truth audio. In contrast, FireRedTTS, which is trained on more high-quality data obtained through the proposed data processing pipeline and employs a larger LM with 400M parameters, achieves better performance with a higher CoMOS score of 4.32. It even surpasses ground-truth audio in some cases. These results underscore the effectiveness of the proposed foundational system.

\begin{table}[htp]
\centering
\begin{tabular}{c|c}
\hline
\multirow{2}{*}{\textbf{System}} & \multirow{2}{*}{\textbf{CoMOS($\uparrow$)}} \\
                                 &                                            \\ \hline
GroundTruth                      & 4.53                                       \\
CosyVoice                        & 4.15                                       \\
FireRedTTS                             & 4.32                                       \\ \hline
\end{tabular}
\caption{The CoMOS evaluation results.}
\label{tab:exp_content_consistency}
\end{table}

\subsubsection{Stability Evaluation}

Next, we evaluate the stability of FireRedTTS by detecting pronunciation errors from 2,000 utterances synthesized from a challenging test set used for product launch testing. Specifically, we select 200 audio prompts with DNSMOS scores ranging from 2.5 to 3.5. For each audio prompt, we generate six Chinese utterances, three English utterances, and one code-switch (Chinese-English) utterance. Professional listeners are employed to detect pronunciation issues in the synthesized audio, providing a more accurate assessment than ASR-based evaluation methods.

Table \ref{tab:exp_content_consistency_detail} presents the sentence-level error rates of CosyVoice and FireRedTTS in three cases: overall, substitution, ``insertion and deletion". The result shows that our system presents better overall performance than CosyVoice, but still has significant stability issues in English and code-switch utterances. Particularly in English, the majority of pronunciation errors are attributable to insertions and deletions. This performance discrepancy can be primarily attributed to the composition of the current training set, which contains a relatively small proportion of English and code-switch data, as well as insufficient diversity in these samples. Addressing these limitations in the training data will be a key focus of our future work to improve the robustness of the system across diverse linguistic contexts.

\begin{table}[htp]
\centering
\begin{tabular}{c|ccc|ccc|ccc}
\hline
\multirow{2}{*}{\textbf{System}} & \multicolumn{3}{c|}{\textbf{Overall(\%)}}                                          & \multicolumn{3}{c|}{\textbf{Sub(\%)}}                                              & \multicolumn{3}{c}{\textbf{\#Ins.\&Del.(\%)}}                                      \\ \cline{2-10} 
                                 & \multicolumn{1}{c|}{\textbf{ZH}} & \multicolumn{1}{c|}{\textbf{EN}} & \textbf{MIX} & \multicolumn{1}{c|}{\textbf{ZH}} & \multicolumn{1}{c|}{\textbf{EN}} & \textbf{MIX} & \multicolumn{1}{c|}{\textbf{ZH}} & \multicolumn{1}{c|}{\textbf{EN}} & \textbf{MIX} \\ \hline
CosyVoice                        & \multicolumn{1}{c|}{5.68}        & \multicolumn{1}{c|}{12.17}        & 29.50        & \multicolumn{1}{c|}{3.76}        & \multicolumn{1}{c|}{6.67}        & 25.00         & \multicolumn{1}{c|}{1.92}        & \multicolumn{1}{c|}{5.50}        & 4.50         \\
Ours                             & \multicolumn{1}{c|}{2.09}        & \multicolumn{1}{c|}{12.00}        & 8.50         & \multicolumn{1}{c|}{1.00}        & \multicolumn{1}{c|}{0.50}        & 4.50         & \multicolumn{1}{c|}{1.09}        & \multicolumn{1}{c|}{11.50}       & 4.00         \\ \hline
\end{tabular}
\caption{The stability evaluation results. ``MIX" denotes the code-switch utterances. ``Sub" denotes the ratio of substitution errors. ``\#Ins.\&Del" denotes the ratio of insertion and deletion errors.}
\label{tab:exp_content_consistency_detail}
\end{table}

\subsubsection{Streamable Decoding}

We compare the synthesis quality of the flow-matching decoder with the proposed streamable decoder via the same CoMOS test introduced in Sec. \ref{ssec:comos}. As presented in Table \ref{tab:exp_decoder}, the non-streaming flow-matching approach demonstrates superior quality, achieving a CoMOS of 4.48. Nonetheless, the streamable decoder emerges as a viable alternative for FireRedTTS, enabling real-time generation capabilities at the cost of a modest reduction in synthesis quality. We notice that the observed quality degradation is partially due to suboptimal Mel spectrogram generation by the Mel codec, which will be improved in our future work.

\begin{table}[htp]
\centering
\begin{tabular}{c|c}
\hline
\multirow{2}{*}{\textbf{System}} & \multirow{2}{*}{\textbf{CoMOS($\uparrow$)}} \\
                                 &                                            \\ \hline
GroundTruth                      & 4.52                                       \\
Flow-matching decoder                        & 4.48                                       \\
Streamable decoder                             & 4.41 \\ \hline
\end{tabular}
\caption{The CoMOS evaluation results of the flow-matching-based decoder and streamable decoder.}
\label{tab:exp_decoder}
\end{table}

\subsection{Voice Cloning}

\subsubsection{Zero-Shot v.s. Few-Shot}


To evaluate the efficacy of voice cloning in dubbing applications, we conduct a comparative analysis of zero-shot and few-shot approaches in both UGC and PUGC scenarios. For the UGC scenario, we randomly select 2 two non-studio unseen speakers (one male, one female) from our internal dataset. The PUGC scenario utilizes a distinctive, highly expressive voice, Wukong, that is underrepresented in our large-scale training set. Our test set comprises 5 audio prompts per speaker, each synthesizing 8 utterances. The zero-shot approach directly uses the foundation system via in-context learning. In contrast, the few-shot approach involves two fine-tuning strategies: (1) For UGC, we fine-tune the TTS language model using 2 minutes of training data; (2) For PUGC, we fine-tune both the TTS language model and the flow-matching decoder using a more extensive 1-hour training set.

We conducted MOS tests and calculated speaker similarity (SIM) using a pre-trained speaker verification model\footnote{\url{https://github.com/modelscope/3D-Speaker}} to evaluate these two approaches both subjectively and objectively. Table~\ref{tab:exp_vc_fewshot_wukong} presents the evaluation results. In the UGC scenario, the zero-shot approach already achieves high-quality synthesis, thanks to the powerful foundational model. Nonetheless, fine-tuning with 2-minute data can further improve performance slightly, which validates the effectiveness of few-shot adaptation in scenarios with extremely limited data. For the PUGC scenario, the zero-shot approach exhibits suboptimal performance in speaker similarity for this distinctive voice. The few-shot approach demonstrates significant improvement, with higher MOS and SIM scores. This result highlights the effectiveness of few-shot fine-tuning in industrial PUGC scenarios.

\begin{table}[!ht]
\centering
\renewcommand{\arraystretch}{1.1}
\begin{tabular}{cccc}
\toprule
Scenario & Approach & \textbf{MOS$(\uparrow)$} & \textbf{SIM(\%)($\uparrow$)} \\ \hline
\multirow{2}{*}{UGC} & zeroshot & 4.25 & 73.61 \\
& fewshot 2min & 4.31 & 73.85 \\ \midrule
\multirow{2}{*}{PUGC} & zeroshot & 3.77 & 68.63 \\
& fewshot 1h & 4.65 & 78.92 \\ \bottomrule
\end{tabular}
\caption{Voice cloning under zero-shot and few-shot conditions for UGC and PUGC scenarios.}
\label{tab:exp_vc_fewshot_wukong}
\end{table}

\subsubsection{Prompt Enhancement}

As discussed in Sec. \ref{ss:clone}, we apply speech enhancement to process noisy audio prompts. To evaluate the effectiveness of this approach, we conduct a similarity MOS test and calculate SIM to audio synthesized with processed or unprocessed audio prompts. We randomly select 40 unseen clean audio prompt (20 in Mandarin and 20 in English), and then manually add 
background noise at various SNR levels, 20 dB, 10 dB, and 0 dB to them, following the method in~\cite{dubey2024icassp}. We then apply the enhancement to the audio prompts and perform the synthesis again. 

Table~\ref{tab:exp_vc_prompt} presents the comparison of synthesized audio with or without prompt enhancement. The findings indicate that preprocessing may hurt voice cloning with high signal-to-noise ratio (SNR) audio prompts. However, with lower-SNR audio prompts, voice cloning performance deteriorates substantially. In these cases, speech enhancement effectively mitigates this degradation. These results suggest that in real-world applications, speech enhancement is an effective strategy for voice cloning, but should be selectively applied only to low-SNR audio prompts.

\begin{table}[!ht]
\centering
\renewcommand{\arraystretch}{1.1}
\begin{tabular}{ccc|cc}
\toprule
\multirow{2}{*}{\textbf{SNR(db)}} & \multicolumn{2}{c|}{\textbf{Similarity MOS}} & \multicolumn{2}{c}{\textbf{SIM(\%)}} \\
 & \textbf{w/o process} & \textbf{with process} & \textbf{w/o process} & \textbf{with process} \\ \hline
20 & 3.97 & 3.89 & 50.66 & 49.20 \\
10 & 3.37 & 3.53 & 41.81 & 43.60 \\
0 & 2.62 & 3.01 & 28.81 & 35.15 \\ \bottomrule
\end{tabular}
\caption{Comparison of voice cloning performance before and after enhancement across three different SNR levels of the audio prompt.}
\label{tab:exp_vc_prompt}
\end{table}

\begin{figure}[htp]
\centering
\includegraphics[width=0.65\linewidth]{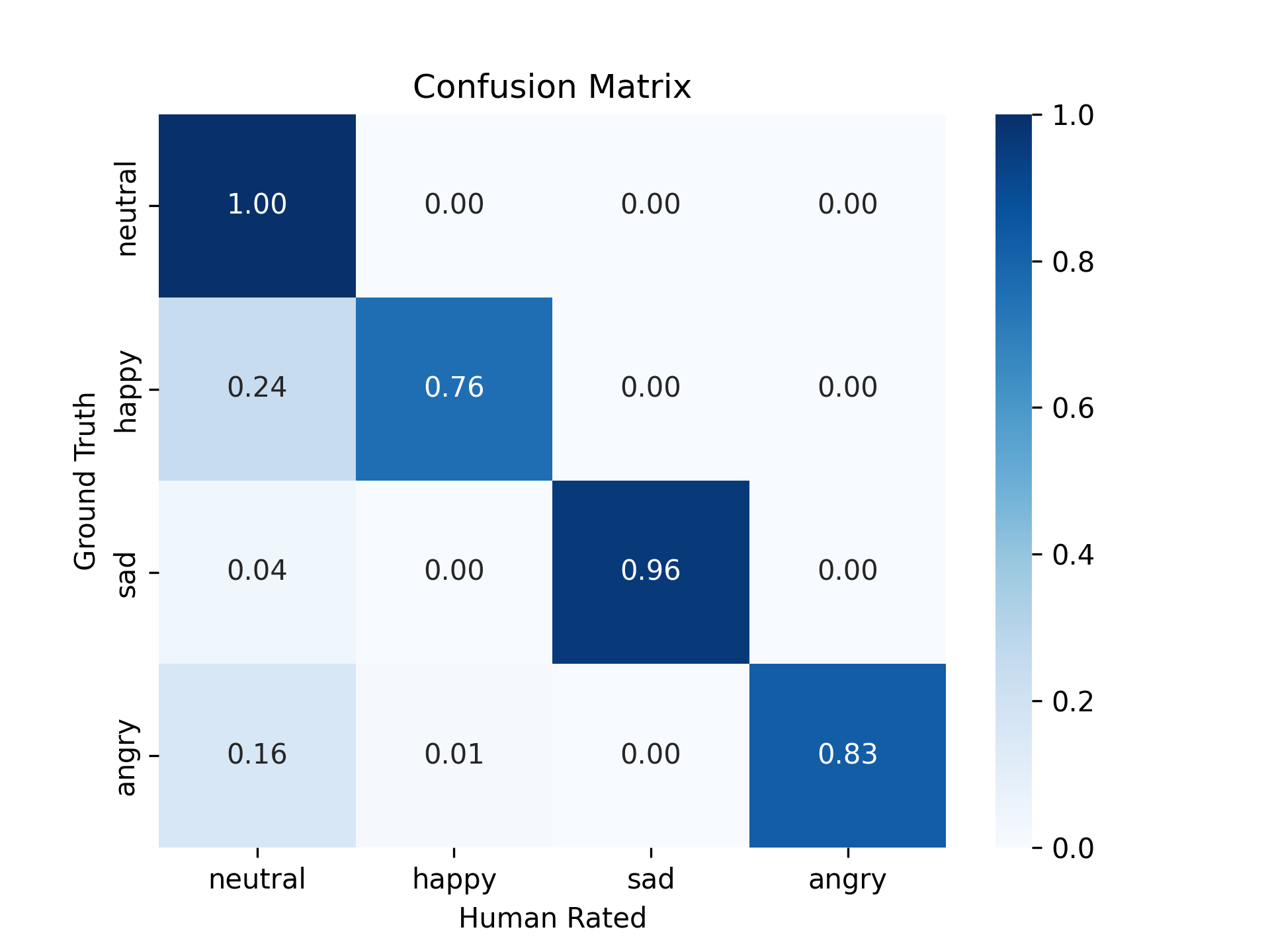}
\caption{The confusion matrix between human rates and ground-truth labels of synthesized emotional speech.}
\label{img:emo_confusion}
\end{figure}

\subsection{Human-like Speech Generation}

To validate the effectiveness of FireRedTTS in human-like speech generation, we evaluate its capability to synthesize emotional speech and paralinguistic behaviors via instruct tuning.

First, we assess the control accuracy of emotional speech synthesis. We construct a single-speaker test set comprising 100 Chinese utterances, each assigned with an emotion in ``neutral'', ``happy,'' ``sad,'' or ``angry''. Listeners are asked to categorize the synthesized audio into one emotion type. As presented in the confusion matrix of Figure~\ref{img:emo_confusion}, the majority of audio samples successfully convey the intended emotion. Only a small subset of samples, characterized by lower emotional intensity, are misclassified as neutral. This result demonstrates that the fine-tuned FireRedTTS is capable of synthesizing emotional speech controllably.

Moreover, to investigate the effect of instruct tuning, we measure the emotional controllability of pre-trained and fine-tuned LMs by calculating the classification accuracy using an emotion recognition model\footnote{\url{https://modelscope.cn/models/iic/emotion2vec_base_finetuned}}. As presented in Table~\ref{tab:exp_emotion_control}, compared with the pre-trained model, instruction tuning significantly improves the model's ability to synthesize speech with the target emotion. This finding further underscores the efficacy of instruction tuning to turn FireRedTTS into an emotion-controllable TTS system.

Finally, we conduct a preference test to assess the impact of paralinguistic behaviors on synthesized speech. As shown in Table~\ref{tab:exp_spon_naturalness}, the incorporation of paralinguistic behaviors significantly improves the human-likeness of the synthesized speech, presenting a higher preference of 45\%. This result underscores the crucial role of paralinguistic behaviors in generating natural and human-like speech for spoken chatbots.

\begin{table}[htp]
\centering
\begin{tabular}{c|cccc}
\toprule
\textbf{Model} & \textbf{Neutral} & \textbf{Happy} & \textbf{Sad} & \textbf{Angry} \\ \hline
pre-trained & 50\% & 45\% & 76\%  & 87\% \\
fine-tuned  & 97\% & 97\% & 100\% & 98\% \\ \bottomrule
\end{tabular}
\caption{The classification accuracy of the emotion recognition model on synthesized speech from the pre-trained and fine-tuned FireRedTTS.}
\label{tab:exp_emotion_control}
\end{table}


\begin{table}[htp]
\centering
\begin{tabular}{ccc}
\toprule
\textbf{w/o} & \textbf{no difference} & \textbf{w/} \\ \hline
26\% & 29\% & 45\%  \\ \bottomrule
\end{tabular}
\caption{The preference test between synthesized speech with or without paralinguistic behaviors.}
\label{tab:exp_spon_naturalness}
\end{table}

\section{Conclusion and Future Work}
This work introduces FireRedTTS, a novel foundation text-to-speech framework. The system comprises three key components: data processing, foundation system, and downstream applications. Initially, we develop a comprehensive data processing pipeline, transforming 624k hours of raw audio into a high-quality, large-scale TTS dataset of 248k hours, encompassing a wide range of content, speaking style, and timbre. Subsequently, we propose a language-model-based foundation TTS system that extracts discrete semantic tokens from speech for text-to-speech modeling and generates high-fidelity audio via a two-stage token-to-waveform generator. We demonstrate two downstream applications of FireRedTTS: voice cloning for dubbing and human-like speech generation for chatbots. Through subjective and objective evaluations, we establish the robust capability of FireRedTTS to generate high-quality speech with expected content, speaking style, and timbre, aligned with the prompt text and audio. Experiments in voice cloning highlight the significant potential of FireRedTTS in dubbing for UGC and PUGC scenarios via zero-shot and few-shot approaches. Furthermore, we showcase FireRedTTS's proficiency in synthesizing emotional speech with paralinguistic behaviors in a controllable manner via instruction tuning, validating its effectiveness in generating human-like speech for chatbots.

\bibliographystyle{unsrt}
\bibliography{refs}

\begin{thebibliography}{10}

\bibitem{guo2021conversational}
Haohan Guo, Shaofei Zhang, Frank~K Soong, Lei He, and Lei Xie.
\newblock {Conversational end-to-end TTS for voice agents}.
\newblock In {\em Proc. SLT}, pages 403--409. IEEE, 2021.

\bibitem{budzianowski2024pheme}
Pawe{\l} Budzianowski, Taras Sereda, Tomasz Cichy, and Ivan Vuli{\'c}.
\newblock Pheme: Efficient and conversational speech generation.
\newblock {\em arXiv preprint arXiv:2401.02839}, 2024.

\bibitem{hu2021neural}
Chenxu Hu, Qiao Tian, Tingle Li, Wang Yuping, Yuxuan Wang, and Hang Zhao.
\newblock {Neural dubber: Dubbing for videos according to scripts}.
\newblock {\em Proc. NeurIPS}, 34, 2021.

\bibitem{liu2024m3tts}
Yan Liu, Li-Fang Wei, Xinyuan Qian, Tian-Hao Zhang, Song-Lu Chen, and Xu-Cheng Yin.
\newblock M3tts: Multi-modal text-to-speech of multi-scale style control for dubbing.
\newblock {\em Pattern Recognition Letters}, 179:158--164, 2024.

\bibitem{touvron2023llama2}
Hugo Touvron, Louis Martin, Kevin Stone, Peter Albert, Amjad Almahairi, Yasmine Babaei, Nikolay Bashlykov, Soumya Batra, Prajjwal Bhargava, Shruti Bhosale, et~al.
\newblock Llama 2: Open foundation and fine-tuned chat models.
\newblock {\em arXiv preprint arXiv:2307.09288}, 2023.

\bibitem{gpt4}
Josh Achiam, Steven Adler, Sandhini Agarwal, Lama Ahmad, Ilge Akkaya, Florencia~Leoni Aleman, Diogo Almeida, Janko Altenschmidt, Sam Altman, Shyamal Anadkat, et~al.
\newblock Gpt-4 technical report.
\newblock {\em arXiv preprint arXiv:2303.08774}, 2023.

\bibitem{gozalo2023chatgpt}
Roberto Gozalo-Brizuela and Eduardo~C Garrido-Merchan.
\newblock Chatgpt is not all you need. a state of the art review of large generative ai models.
\newblock {\em arXiv preprint arXiv:2301.04655}, 2023.

\bibitem{VALLEX}
Ziqiang Zhang, Long Zhou, Chengyi Wang, Sanyuan Chen, Yu~Wu, Shujie Liu, Zhuo Chen, Yanqing Liu, Huaming Wang, Jinyu Li, Lei He, Sheng Zhao, and Furu Wei.
\newblock Speak foreign languages with your own voice: Cross-lingual neural codec language modeling.
\newblock {\em CoRR}, abs/2303.03926, 2023.

\bibitem{tortoise}
James Betker.
\newblock Better speech synthesis through scaling.
\newblock {\em arXiv preprint arXiv:2305.07243}, 2023.

\bibitem{lajszczak2024base}
Mateusz {\L}ajszczak, Guillermo C{\'a}mbara, Yang Li, Fatih Beyhan, Arent van Korlaar, Fan Yang, Arnaud Joly, {\'A}lvaro Mart{\'\i}n-Cortinas, Ammar Abbas, Adam Michalski, et~al.
\newblock Base tts: Lessons from building a billion-parameter text-to-speech model on 100k hours of data.
\newblock {\em arXiv preprint arXiv:2402.08093}, 2024.

\bibitem{anastassiou2024seed}
Philip Anastassiou, Jiawei Chen, Jitong Chen, Yuanzhe Chen, Zhuo Chen, Ziyi Chen, Jian Cong, Lelai Deng, Chuang Ding, Lu~Gao, et~al.
\newblock Seed-tts: A family of high-quality versatile speech generation models.
\newblock {\em arXiv preprint arXiv:2406.02430}, 2024.

\bibitem{du2024cosyvoice}
Zhihao Du, Qian Chen, Shiliang Zhang, Kai Hu, Heng Lu, Yexin Yang, Hangrui Hu, Siqi Zheng, Yue Gu, Ziyang Ma, et~al.
\newblock Cosyvoice: A scalable multilingual zero-shot text-to-speech synthesizer based on supervised semantic tokens.
\newblock {\em arXiv preprint arXiv:2407.05407}, 2024.

\bibitem{dong2022survey}
Qingxiu Dong, Lei Li, Damai Dai, Ce~Zheng, Zhiyong Wu, Baobao Chang, Xu~Sun, Jingjing Xu, and Zhifang Sui.
\newblock A survey on in-context learning.
\newblock {\em arXiv preprint arXiv:2301.00234}, 2022.

\bibitem{tacotron2}
Jonathan Shen, Ruoming Pang, Ron~J. Weiss, Mike Schuster, Navdeep Jaitly, Zongheng Yang, Zhifeng Chen, Yu~Zhang, Yuxuan Wang, R.~J. Skerry{-}Ryan, Rif~A. Saurous, Yannis Agiomyrgiannakis, and Yonghui Wu.
\newblock Natural {TTS} synthesis by conditioning wavenet on mel spectrogram predictions.
\newblock {\em CoRR}, abs/1712.05884, 2017.

\bibitem{fastspeech2}
Yi~Ren, Chenxu Hu, Xu~Tan, Tao Qin, Sheng Zhao, Zhou Zhao, and Tie{-}Yan Liu.
\newblock {FastSpeech 2: Fast and high-quality end-to-end text to speech}.
\newblock In {\em Proc. ICLR}, 2021.

\bibitem{rouard2023hybrid}
Simon Rouard, Francisco Massa, and Alexandre D{\'e}fossez.
\newblock Hybrid transformers for music source separation.
\newblock In {\em Proc. ICASSP}, pages 1--5. IEEE, 2023.

\bibitem{schroter2022deepfilternet}
Hendrik Schroter, Alberto~N Escalante-B, Tobias Rosenkranz, and Andreas Maier.
\newblock Deepfilternet: A low complexity speech enhancement framework for full-band audio based on deep filtering.
\newblock In {\em Proc. ICASSP}, pages 7407--7411. IEEE, 2022.

\bibitem{bai2019voice}
Ye~Bai, Jiangyan Yi, Jianhua Tao, Zhengqi Wen, and Bin Liu.
\newblock Voice activity detection based on time-delay neural networks.
\newblock In {\em 2019 Asia-Pacific Signal and Information Processing Association Annual Summit and Conference (APSIPA ASC)}, pages 1173--1178. IEEE, 2019.

\bibitem{wang2023wespeaker}
Hongji Wang, Chengdong Liang, Shuai Wang, Zhengyang Chen, Binbin Zhang, Xu~Xiang, Yanlei Deng, and Yanmin Qian.
\newblock Wespeaker: A research and production oriented speaker embedding learning toolkit.
\newblock In {\em Proc. ICASSP}, pages 1--5. IEEE, 2023.

\bibitem{sainath2019two}
Tara~N Sainath, Ruoming Pang, David Rybach, Yanzhang He, Rohit Prabhavalkar, Wei Li, Mirk{\'o} Visontai, Qiao Liang, Trevor Strohman, Yonghui Wu, et~al.
\newblock Two-pass end-to-end speech recognition.
\newblock {\em arXiv preprint arXiv:1908.10992}, 2019.

\bibitem{reddy2022dnsmos}
Chandan~KA Reddy, Vishak Gopal, and Ross Cutler.
\newblock Dnsmos p. 835: A non-intrusive perceptual objective speech quality metric to evaluate noise suppressors.
\newblock In {\em Proc. ICASSP}, pages 886--890. IEEE, 2022.

\bibitem{yu2024autoprep}
Jianwei Yu, Hangting Chen, Yanyao Bian, Xiang Li, Yi~Luo, Jinchuan Tian, Mengyang Liu, Jiayi Jiang, and Shuai Wang.
\newblock Autoprep: An automatic preprocessing framework for in-the-wild speech data.
\newblock In {\em Proc. ICASSP}, pages 1136--1140. IEEE, 2024.

\bibitem{hsu2021hubert}
Wei-Ning Hsu, Benjamin Bolte, Yao-Hung~Hubert Tsai, Kushal Lakhotia, Ruslan Salakhutdinov, and Abdelrahman Mohamed.
\newblock Hubert: Self-supervised speech representation learning by masked prediction of hidden units.
\newblock {\em IEEE/ACM Transactions on Audio, Speech, and Language Processing}, 29:3451--3460, 2021.

\bibitem{guo2024addressing}
Haohan Guo, Fenglong Xie, Dongchao Yang, Hui Lu, Xixin Wu, and Helen Meng.
\newblock Addressing index collapse of large-codebook speech tokenizer with dual-decoding product-quantized variational auto-encoder.
\newblock {\em arXiv preprint arXiv:2406.02940}, 2024.

\bibitem{dawalatabad2021ecapa}
Nauman Dawalatabad, Mirco Ravanelli, Fran{\c{c}}ois Grondin, Jenthe Thienpondt, Brecht Desplanques, and Hwidong Na.
\newblock Ecapa-tdnn embeddings for speaker diarization.
\newblock {\em arXiv preprint arXiv:2104.01466}, 2021.

\bibitem{vqvae}
Aaron van~den Oord, Oriol Vinyals, and koray kavukcuoglu.
\newblock Neural discrete representation learning.
\newblock In I.~Guyon, U.~Von Luxburg, S.~Bengio, H.~Wallach, R.~Fergus, S.~Vishwanathan, and R.~Garnett, editors, {\em Proc. NeurIPS}, volume~30. Curran Associates, Inc., 2017.

\bibitem{whisper}
Alec Radford, Jong~Wook Kim, Tao Xu, Greg Brockman, Christine McLeavey, and Ilya Sutskever.
\newblock Robust speech recognition via large-scale weak supervision.
\newblock In {\em Proc. ICML}, pages 28492--28518. PMLR, 2023.

\bibitem{brown2020language}
Tom Brown, Benjamin Mann, Nick Ryder, Melanie Subbiah, Jared~D Kaplan, Prafulla Dhariwal, Arvind Neelakantan, Pranav Shyam, Girish Sastry, Amanda Askell, et~al.
\newblock Language models are few-shot learners.
\newblock {\em Proc. NeurIPS}, 33:1877--1901, 2020.

\bibitem{liu2024audiosr}
Haohe Liu, Ke~Chen, Qiao Tian, Wenwu Wang, and Mark~D Plumbley.
\newblock Audiosr: Versatile audio super-resolution at scale.
\newblock In {\em Proc. ICASSP}, pages 1076--1080. IEEE, 2024.

\bibitem{mehta2024matcha}
Shivam Mehta, Ruibo Tu, Jonas Beskow, {\'E}va Sz{\'e}kely, and Gustav~Eje Henter.
\newblock Matcha-tts: A fast tts architecture with conditional flow matching.
\newblock In {\em Proc. ICASSP}, pages 11341--11345. IEEE, 2024.

\bibitem{lipman2022flow}
Yaron Lipman, Ricky~TQ Chen, Heli Ben-Hamu, Maximilian Nickel, and Matt Le.
\newblock Flow matching for generative modeling.
\newblock {\em arXiv preprint arXiv:2210.02747}, 2022.

\bibitem{ho2022classifier}
Jonathan Ho and Tim Salimans.
\newblock Classifier-free diffusion guidance.
\newblock {\em arXiv preprint arXiv:2207.12598}, 2022.

\bibitem{socodec}
Haohan Guo, Fenglong Xie, Kun Xie, Dongchao Yang, Dake Guo, Xixin Wu, and Helen Meng.
\newblock Socodec: A semantic-ordered multi-stream speech codec for efficient language model based text-to-speech synthesis.
\newblock {\em arXiv preprint arXiv:2409.00933}, 2024.

\bibitem{copet2024simple}
Jade Copet, Felix Kreuk, Itai Gat, Tal Remez, David Kant, Gabriel Synnaeve, Yossi Adi, and Alexandre D{\'e}fossez.
\newblock Simple and controllable music generation.
\newblock {\em Proc. NeurIPS}, 36, 2024.

\bibitem{bigvgan}
{Sang-gil} Lee, Wei Ping, Boris Ginsburg, Bryan Catanzaro, and Sungroh Yoon.
\newblock Big{VGAN}: A universal neural vocoder with large-scale training.
\newblock In {\em Proc. ICLR}, 2023.

\bibitem{dubey2024icassp}
Harishchandra Dubey, Ashkan Aazami, Vishak Gopal, Babak Naderi, Sebastian Braun, Ross Cutler, Alex Ju, Mehdi Zohourian, Min Tang, Mehrsa Golestaneh, et~al.
\newblock Icassp 2023 deep noise suppression challenge.
\newblock {\em IEEE Open Journal of Signal Processing}, 2024.

\end{thebibliography}


\end{document}